\documentclass[sigconf]{acmart}

\usepackage{multirow}
\usepackage{balance}
\AtBeginDocument{%
  \providecommand\BibTeX{{%
    \normalfont B\kern-0.5em{\scshape i\kern-0.25em b}\kern-0.8em\TeX}}}
\copyrightyear{2022}
\acmYear{2022}
\setcopyright{acmcopyright}
\acmConference[CIKM '22] {Proceedings of the 31st ACM International Conference on Information and Knowledge Management}{October 17--21, 2022}{Atlanta, GA, USA}
\acmBooktitle{Proceedings of the 31st ACM International Conference on Information and Knowledge Management (CIKM '22), October 17--21, 2022, Atlanta, GA, USA}
\acmPrice{15.00}
\acmISBN{978-1-4503-9236-5/22/10}
\acmDOI{10.1145/3511808.3557639}

\begin{document}

\title{META-CODE: Community Detection via Exploratory Learning in Topologically Unknown Networks}


\author{Yu Hou}
\orcid{1234-5678-9012}
\affiliation{%
  \institution{Yonsei University}
  \streetaddress{P.O. Box 1212}
  \city{Seoul}
  \country{Republic of Korea}
  \postcode{43017-6221}
}
\email{houyu@yonsei.ac.kr}

\author{Cong Tran}
\affiliation{%
  \institution{Posts and Telecommunications Institute of Technology}
  \city{Hanoi}
  \country{Vietnam}}
\email{congtt@ptit.edu.vn}

\author{Won-Yong Shin}
\affiliation{%
  \institution{Yonsei University}
  \city{Seoul}
  \country{Republic of Korea}
}
\email{wy.shin@yonsei.ac.kr}

\authornote{Corresponding author}

\renewcommand{\shortauthors}{Yu Hou, Cong Tran, \& Won-Yong Shin}

\begin{abstract}
  The discovery of community structures in social networks has gained considerable attention as a fundamental problem for various network analysis tasks. However, due to privacy concerns or access restrictions, the network structure is often {\em unknown}, thereby rendering established community detection approaches ineffective without costly data acquisition. To tackle this challenge, we present \textsf{META-CODE}, a novel end-to-end solution for detecting overlapping communities in networks with unknown topology via {\em exploratory learning} aided by {\it easy-to-collect} node metadata. Specifically, \textsf{META-CODE} consists of three steps: 1) initial network inference, 2) node-level community-affiliation embedding based on graph neural networks (GNNs) trained by our new reconstruction loss, and 3) network exploration via community-affiliation-based node queries, where Steps 2 and 3 are performed iteratively. Experimental results demonstrate that \textsf{META-CODE} exhibits (a) superiority over benchmark methods for overlapping community detection, (b) the effectiveness of our training model, and (c) fast network exploration.
\end{abstract}





\maketitle

\section{Introduction}
Community detection \cite{lancichinetti2009community, papadopoulos2012community, xie2013overlapping, cao2018behavior, kang2020cr, ding2021deep} is an indispensable task in network analyses to understand the fundamental features of networks. Community detection algorithms should be designed by taking into account the inherently overlapping nature since community memberships in social networks are allowed to overlap as nodes belong to multiple clusters at once \cite{shchur2019overlapping, yang2012community, yang2013overlapping, yang2013community, ye2018deep, tran2021community}.

Nevertheless, the underlying {\em true} network is not initially known in various domains of real-world network applications \cite{valente2007identifying}. Thus, existing downstream graph mining approaches are often ineffective without retrieving relationships among nodes in a network, which is very labor-intensive. To overcome this problem, {\em exploratory} influence maximization was introduced along with a solution to perform information retrieval over unknown networks by querying individual nodes to explore their neighborhoods, while aiming at discovering the set of seed nodes leading to the maximum influence spread \cite{wilder2018maximizing, wilder2018end, kamarthi2019influence, tran2021meta}. However, the aforementioned studies focused on solving the problem of influence maximization; the research related to community detection via exploratory learning has been largely underexplored yet.

In social networks, node metadata can often be retrieved from a variety of sources such as user-created content over given (social) networks \cite{leroy2010cold} or even be inferred from users' behavior \cite{kosinski2013private}, thereby making them more readily available than the relationship between individuals. In this paper, motivated by the fact that collecting such node metadata is relatively {\it easy-to-collect} and cost-effective \cite{leroy2010cold}, we present \textsf{META-CODE}, a novel end-to-end framework for overlapping community detection in networks with unknown topology via {\em exploratory learning} that retrieves information from both node queries and node metadata. As the first step of \textsf{META-CODE}, we infer the network structure by using both the multi-assignment clustering (MAC) \cite{streich2009multi} and the {\it community-affiliation graph model (AGM)} \cite{yang2012community}. Then, \textsf{META-CODE} performs the following two steps {\em iteratively}: we generate node-level community-affiliation embeddings based on graph neural networks (GNNs) trained by our new loss function, which is established for preserving structural and attribute information; and, for network exploration, we identify the next node to query based on the discovered community affiliations.

\begin{figure}[htbp]
\begin{center}
    \includegraphics[width=1.0\linewidth]{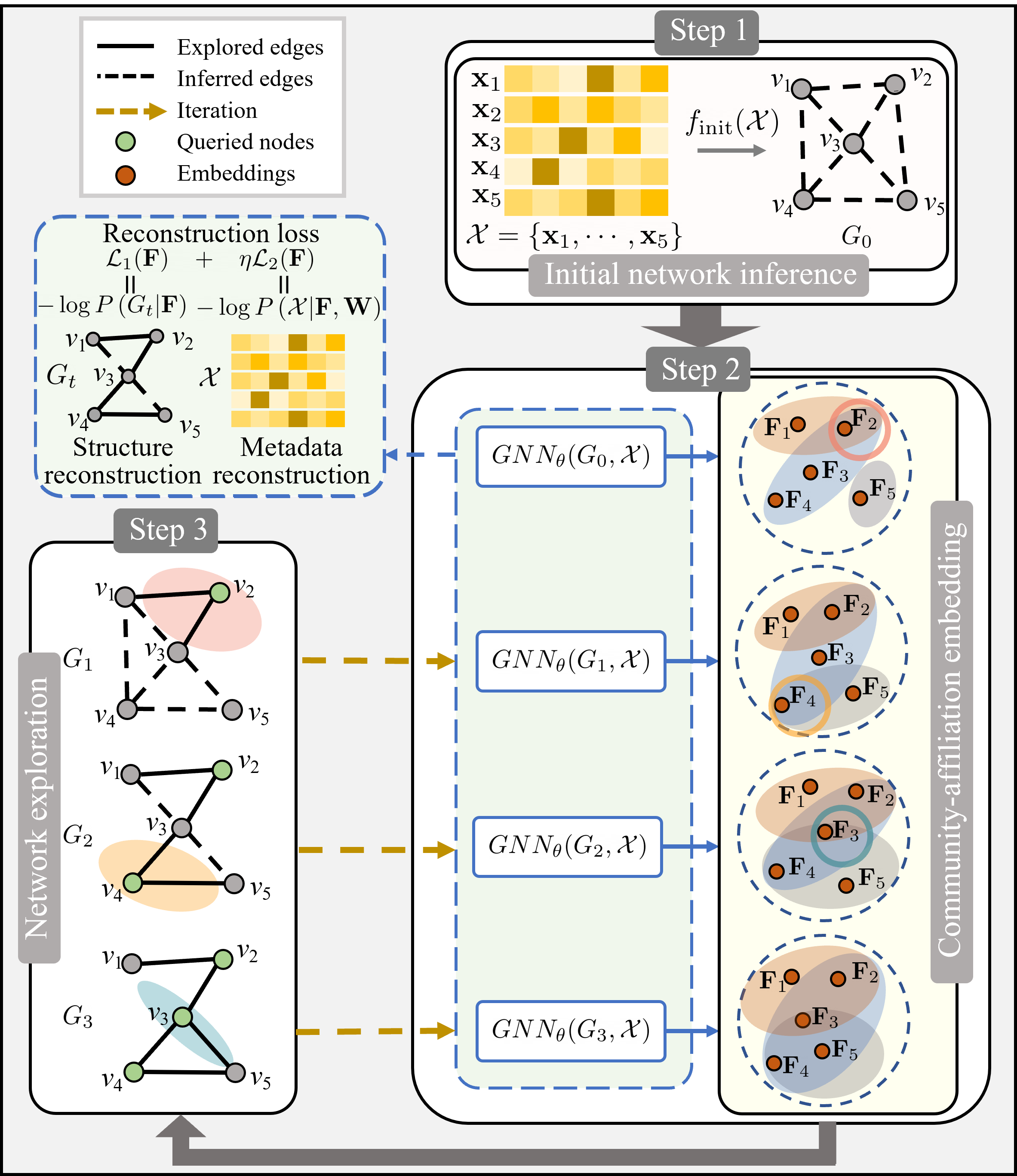}
\end{center}
   \caption{The schematic overview of \textsf{META-CODE} when a query budget $T$ is given by 3.}
\label{fig:overview}
\end{figure}

Through comprehensive experiments using real-world datasets, we demonstrate that \textsf{META-CODE} consistently outperforms benchmark methods for overlapping community detection. We also validate the effectiveness of both our network inference module and our training model in GNNs. Moreover, the proposed node query strategy in \textsf{META-CODE} shows faster network exploration than competing sampling strategies, thus potentially leading to superior performance on community detection.

\section{Methodology}

\subsection{Basic Settings and Problem Definition}

Let us denote an underlying network as $G = \left( {\mathcal{V},\mathcal{E}} \right)$, which is initially unavailable, where $\mathcal{V}$ is the set of $N$ nodes and $\mathcal{E}$ is the set of edges. We assume $G$ to be an undirected unweighted {\em attributed} network having collectible node metadata ${\bf \mathcal{X}} \in \mathbb{R}^{N \times D}$ of binary-valued node features\footnote{Non binary-valued node features can be transferred to binary values via one-hot encoding \cite{leskovec2012learning}.}, where $D$ is the dimension of each feature vector. We assume that social networks follow the AGM \cite{yang2012community}, which can be represented by a non-negative weight affiliation matrix ${\bf F}\in\mathbb{R}^{N\times C}$ regarded as node-level community-affiliation embeddings, where $C$ is the number of communities and ${\bf F}_u$ is the affiliations of node $u$. Following the concept of active learning [2,12], we are allowed to perform $T$ node queries given a query budget $T$ for network exploration. When we query a single node, we are capable of discovering neighbors of the queried node. The observable subgraph can then be expanded accordingly, which essentially follows the same setting as prior work \cite{wilder2018end, wilder2018maximizing, mihara2015influence}.

\textbf{Problem definition:} Given an initially unknown network $G = \left( {\mathcal{V},\mathcal{E}} \right)$, node metadata $\mathcal{X}$, and $T$ node queries, we aim to assign nodes to $C$ overlapping communities.
\subsection{Methodology}

\subsubsection{Overview of META-CODE}

We describe our \textsf{META-CODE} method composed of three steps. The schematic overview of \textsf{META-CODE} for $T=3$ is illustrated in Figure \ref{fig:overview}.

In Step 1, we discover the initial communities using the MAC \cite{streich2009multi} that assigns nodes to multiple communities based on the node metadata $\mathcal{X}$.  We then infer the network, denoted as $G_0=(\mathcal{V},\mathcal{E}_0)$, from the initially discovered communities using the AGM, where $\mathcal{E}_0$ is the set of initially inferred edges. The function of initial network inference is denoted as $G_0  = f_\text{init}(\mathcal{X})$. In Step 2, we generate node-level community-affiliation embeddings by designing a new GNN model. Specifically, we build a GNN architecture and then train our GNN model by establishing our own reconstruction loss in terms of preserving structural and attribute information. In Step 3, we explore a subgraph by selecting an influential node through the query process. To this end, we present our query node selection strategy, which is inspired by the claim that querying nodes not only belonging to more communities but also being distributed over diverse communities is more influential in enabling the explored subgraph to grow faster, thus resulting in potentially better performance on community detection. This claim will be empirically validated in Section 3. In Sections 2.2.2 and 2.2.3, we elaborate on the embedding and network exploration steps, respectively.

\subsubsection{Community-Affiliation Embedding via GNNs}

Let $G_t=(\mathcal{V},\mathcal{E}_t)$ denote the discovered network whose explored parts are expanded by replacing the connectivity information with some amount of uncertainty for a queried node at the $t$-th query step ($t\in\{1,\cdots,T\}$) by connections between the queried node and its neighbors, as depicted in Figure 1. Using the discovered network $G_t$ and node metadata $\mathcal{X}$, we aim at designing our GNN model to generate node-level embeddings that are represented as a non-negative weight affiliation matrix ${\bf F}$:
\begin{align}
{\bf F}:= GNN_\theta (G_t,\mathcal{X}).
\end{align}

Toward this goal, we would like to characterize the following two relationships.

\textbf{Structure-community relationship:} Given the affiliation matrix ${\bf F}$, the probability of having an edge between two nodes $u$ and $v$ can be modeled as ${1 - \exp \left( { - {\bf F}_u {\bf F}_v^T } \right)}$, which is built upon the probabilistic generative model in \cite{yang2013overlapping}. This implies that, if a pair of nodes belongs to more communities in common, then the probability that the node pair is connected becomes higher. In this context, the negative log-likelihood (NLL) can be formulated as $\mathcal{L}_1 \left( {\bf F} \right) =  - \log P\left( {G_t |{\bf F}} \right)$.

\textbf{Metadata-community relationship:} Intuitively, there exists a relationship between node metadata and nodes' affiliations. According to attribute modeling in \cite{yang2013community}, the $d$-th attribute $\mathcal{X}_{ud}$ of node $u$ can be probabilistically modeled as
\begin{equation}
Q_{ud}  = \frac{1 }{{1 + \exp \left( { - \sum\nolimits_c {W_{dc}  \cdot {\bf F}_{uc} } } \right)}},
\end{equation}
where ${\bf F}_{uc}$ is the $c$-th entry in ${\bf F}_u$, and $W_{dc}$, the $(d,c)$-th entry of weight matrix ${\bf W}\in\mathbb{R}^{D\times C}$, indicates the relevance of community $c$ to the $d$-th node feature. Thus, the NLL can be formulated as $\mathcal{L}_2 \left( {\bf F} \right) =  - \log P\left( {\mathcal{X} |{\bf F},{\bf W}} \right)$.

\textbf{Model training:} Based on the two relationships above, we establish our reconstruction loss as follows:
\begin{equation}
\begin{aligned}
\mathcal{L}\left ( {\bf F} \right )&=\mathcal{L}_1({\bf F})+\eta \mathcal{L}_2({\bf F}) \\
  &= - \sum\limits_{\left( {u,v} \right) \in \mathcal{E}_t} {\log \left( {1 - \exp \left( { - {\bf F}_u {\bf F}_v^T } \right)} \right) + \sum\limits_{\left( {u,v} \right) \notin \mathcal{E}_t} {{\bf F}_u {\bf F}_v^T } } \\
\end{aligned}
\notag
\end{equation}
\begin{equation}
\;\;\;\;\;\;\;- \eta \sum\limits_{u,d} {\left( {\mathcal{X}_{ud} \log Q_{ud}  + \left( {1 - \mathcal{X}_{ud} } \right)\log \left( {1 - Q_{ud} } \right)} \right)},
 \label{eq:loss}
\end{equation}
where $\eta \ge0$ is the hyperparameter to balance between two terms. Note that the first and second terms in Eq. (\ref{eq:loss}) aim at reconstructing the network structure and node metadata, respectively. We update the parameters $\theta$ of our GNN model in the sense of minimizing the loss function.

\subsubsection{Network Exploration via Node Queries}

As stated in Section 2.1, we discover neighbors of each queried node for network exploration. Thus, we are interested in how to select the set of $T$ queried nodes. Although querying high-degree nodes is expected to make the explored subgraph grow fastest, it may not be true as long as the community detection task is concerned; moreover, such topology-aware query strategies would not be feasible due to the {\it neighbor unawareness} of nodes (e.g., the degree of each node) beforehand. In our study, the node query strategy is built upon our claim that each queried node should not only belong to more communities but also be distributed over diverse communities for faster network exploration. In this context, we select the $t$-th node to query for $t\in\{1,\cdots,T\}$ according to the following formulation:


\begin{equation}
\hat{u}_t = \mathop {\arg \max }\limits_u \left( {\left\| {F_u } \right\|_1  + \lambda\left( {1 - \frac{1}{{\left| \mathcal{P}_t \right|}}\sum\limits_{v \in \mathcal{P}_t} {\text{sim}\left( {F_u ,F_v } \right)} } \right)} \right),
\label{eq:strategy}
\end{equation}
where $\left\|  \cdot  \right\|_1$ is $L_1$-norm of a vector, $\mathcal{P}_t$ is the set of queried nodes up to the $t$-th iteration, $\lambda \ge 0$ is the hyperparameter to balance between two terms, and $\text{sim}(\cdot,\cdot)$ is the cosine similarity. Here, the second term in Eq. (4) controls a node to be selected from communities different from those of previous queried nodes.





\section{Experimental Evaluation}

\textbf{Datasets.} We conduct our experiments on two Facebook datasets \cite{leskovec2012learning}, which are ego-networks collected from Facebook social networks, and one large network dataset, Engineering, which is a co-authorship network constructed from the Microsoft Academic Graph \cite{mac} where communities correspond to research areas and node metadata are based on keywords of the papers by each author\footnote{Although there are other Facebook and large co-authorship networks, we have not further adopted them since they exhibit similar tendencies to those of the above three datasets under consideration.}. 

\textbf{Performance metrics.} We quantify the degree of agreement between the ground truth communities and the detected communities by adopting the normalized mutual information (NMI) and the $F_1$ score. We note that the two metrics are in a range of $[0,1]$, and higher values represent better performance.

\begin{table*}[ht]
\setlength\tabcolsep{5.0pt}
\small
  \caption{Performance comparison of \textsf{META-CODE} and six competing methods in terms of two metrics (average $\pm$ standard deviation) when different portions (\%) of nodes are queried among $N$ nodes. Here, the best and second best performers are highlighted by bold and underline, respectively.}
  \label{tab:comparison}
  \begin{tabular}{cc|cc|cc|cc|ccl}
    \toprule[1pt]
    \multicolumn{2}{c|}{}&\multicolumn{2}{|c|}{10\% queried nodes}&\multicolumn{2}{c|}{20\% queried nodes}&\multicolumn{2}{c|}{30\% queried nodes}&\multicolumn{2}{c}{40\% queried nodes}\\
    \cmidrule{1-10}
           Dataset&Method & NMI& $F_1$ score& NMI& $F_1$ score& NMI& $F_1$ score& NMI& $F_1$ score\\
    \midrule[1pt]
    \multirow{7}*{\rotatebox{90}{Facebook 0}} &
    BIGCLAM+RS & 0.0058\tiny$\pm$0.0012& 0.4048\tiny$\pm$\tiny0.0092& 0.0189\tiny$\pm$0.0088& 0.4253\tiny$\pm$0.0156& 0.0252\tiny$\pm$0.0094& 0.4535\tiny$\pm$0.0056& 0.0348\tiny$\pm$0.0095& 0.5168\tiny$\pm$0.0046\\
    & BIGCLAM+DFS& 0.0206\tiny$\pm$0.0092& 0.4131\tiny$\pm$0.0108& 0.0218\tiny$\pm$0.0047& 0.4217\tiny$\pm$0.0426& 0.0242\tiny$\pm$0.0095& 0.4514\tiny$\pm$0.0147& 0.0263\tiny$\pm$0.0112 & 0.4576\tiny$\pm$0.0652\\
    & vGraph+RS& 0.0079\tiny$\pm$0.0036& 0.4102\tiny$\pm$0.0681& 0.0195\tiny$\pm$0.0130& 0.4989\tiny$\pm$0.0482& 0.0217\tiny$\pm$0.0058& 0.5026\tiny$\pm$0.0233& 0.0410\tiny$\pm$0.0131 & 0.5429\tiny$\pm$0.0480\\
    & vGraph+DFS& 0.0180\tiny$\pm$0.0070& 0.4826\tiny$\pm$0.0269& 0.0301\tiny$\pm$0.0077& 0.5029\tiny$\pm$0.0147& 0.0321\tiny$\pm$0.0116& 0.5066\tiny$\pm$0.0237& 0.0335\tiny$\pm$0.0035& 0.5173\tiny$\pm$0.0186\\
    & NOCD+RS& \underline{0.0256}\tiny$\pm$0.0312& \underline{0.5052}\tiny$\pm$0.0431& \underline{0.0328}\tiny$\pm$0.0301& 0.5319\tiny$\pm$0.0216& \underline{0.0447}\tiny$\pm$0.0234& \underline{0.5539}\tiny$\pm$0.0676& \underline{0.0450}\tiny$\pm$0.0157& \underline{0.6047}\tiny$\pm$0.0395\\
    & NOCD+DFS& 0.0198\tiny$\pm$0.0057& 0.5018\tiny$\pm$0.0504& 0.0290\tiny$\pm$0.0029& \underline{0.5467}\tiny$\pm$0.0128& 0.0308\tiny$\pm$0.0216& 0.5506\tiny$\pm$0.0294& 0.0391\tiny$\pm$0.0243& 0.5938\tiny$\pm$0.0551\\
    & \textsf{META-CODE}& {\bf 0.0298}\tiny$\pm$0.0157& {\bf 0.5119}\tiny$\pm$0.0432& {\bf 0.0357}\tiny$\pm$0.0348& {\bf 0.5596}\tiny$\pm$0.0313& {\bf 0.0463}\tiny$\pm$0.0155& {\bf 0.5623}\tiny$\pm$0.0484& {\bf 0.0556}\tiny$\pm$0.0119& {\bf 0.6102}\tiny$\pm$0.0433\\
    \midrule[1pt]
    \multirow{7}*{\rotatebox{90}{Facebook 348}} &
    BIGCLAM+RS & 0.0263\tiny$\pm$0.0056& 0.4349\tiny$\pm$0.0213& 0.0539\tiny$\pm$0.0193& 0.5025\tiny$\pm$0.0681& 0.0703\tiny$\pm$0.0135& 0.5896\tiny$\pm$0.0117& 0.1083\tiny$\pm$0.0162& 0.6503\tiny$\pm$0.1299\\
    & BIGCLAM+DFS& 0.0379\tiny$\pm$0.0057& 0.4757\tiny$\pm$0.0115& 0.0571\tiny$\pm$0.0211& 0.5364\tiny$\pm$0.0351& 0.0657\tiny$\pm$0.0197& 0.5452\tiny$\pm$0.0733& 0.0785\tiny$\pm$0.0139& 0.5987\tiny$\pm$0.0448\\
    & vGraph+RS& 0.0559\tiny$\pm$0.0152& 0.5700\tiny$\pm$0.0149& 0.0914\tiny$\pm$0.0089& \underline{0.7046}\tiny$\pm$0.0550& 0.1332\tiny$\pm$0.0116& 0.7233\tiny$\pm$0.0461& 0.1573\tiny$\pm$0.0098& 0.7708\tiny$\pm$0.0398\\
    & vGraph+DFS& 0.0601\tiny$\pm$0.0221& 0.5526\tiny$\pm$0.0527& 0.0943\tiny$\pm$0.0104& 0.6614\tiny$\pm$0.0288& 0.1030\tiny$\pm$0.0180& 0.7079\tiny$\pm$0.0584& 0.1272\tiny$\pm$0.0103& 0.7297\tiny$\pm$0.0356\\
    & NOCD+RS& \underline{0.0704}\tiny$\pm$0.0191& \underline{0.6503}\tiny$\pm$0.0960& \underline{0.1022}\tiny$\pm$0.0089& 0.6627\tiny$\pm$0.0836& \underline{0.1422}\tiny$\pm$0.0111& \underline{0.7445}\tiny$\pm$0.0168& \underline{0.1958}\tiny$\pm$0.0191& 0.7652\tiny$\pm$0.0399\\
    & NOCD+DFS& 0.0594\tiny$\pm$0.0111& 0.5471\tiny$\pm$0.0283& 0.0965\tiny$\pm$0.0213& 0.6568\tiny$\pm$0.0616& 0.1329\tiny$\pm$0.0218& 0.7352\tiny$\pm$0.0713& 0.1683\tiny$\pm$0.0247& \underline{0.7885}\tiny$\pm$0.0542\\
    & \textsf{META-CODE}& {\bf 0.0791}\tiny$\pm$0.0198& {\bf 0.6842}\tiny$\pm$0.0343& {\bf 0.1132}\tiny$\pm$0.0167& {\bf 0.7221}\tiny$\pm$0.0567& {\bf 0.1578}\tiny$\pm$0.0169& {\bf 0.7923}\tiny$\pm$0.0493& {\bf 0.2003}\tiny$\pm$0.0164& {\bf 0.8218}\tiny$\pm$0.0085\\
    \midrule[1pt]
    \multirow{7}*{\rotatebox{90}{Engineering}} &
    BIGCLAM+RS & 0.0000\tiny$\pm$0.0000& 0.1315\tiny$\pm$0.0386 & 0.0000\tiny$\pm$0.0000& 0.2635\tiny$\pm$0.0207 &0.0000\tiny$\pm$0.0000 &0.3015\tiny$\pm$0.0531 &0.0144\tiny$\pm$0.0133 & 0.4659\tiny$\pm$0.0317\\
    & BIGCLAM+DFS& 0.0029\tiny$\pm$0.0066& 0.3460\tiny$\pm$0.0370& 0.0194\tiny$\pm$0.0123& 0.4530\tiny$\pm$0.0674& 0.0406\tiny$\pm$0.0192& 0.5160\tiny$\pm$0.0351& 0.0590\tiny$\pm$0.0088& 0.5980\tiny$\pm$0.0387\\
    & vGraph+RS& 0.0000\tiny$\pm$0.0000& 0.1972\tiny$\pm$0.0282& 0.0000\tiny$\pm$0.0000& 0.3334\tiny$\pm$0.0080& 0.0334\tiny$\pm$0.0042& 0.4702\tiny$\pm$0.0188& 0.0522\tiny$\pm$0.0116& 0.5877\tiny$\pm$0.0913\\
    & vGraph+DFS& 0.0000\tiny$\pm$0.0000& 0.3219\tiny$\pm$0.0084& 0.0281\tiny$\pm$0.0114& 0.4449\tiny$\pm$0.0642& 0.0368\tiny$\pm$0.0054& 0.4811\tiny$\pm$0.0820& 0.0504\tiny$\pm$0.0194& 0.5217\tiny$\pm$0.0137\\
    & NOCD+RS& 0.1262\tiny$\pm$0.0219& 0.7081\tiny$\pm$0.0277& \underline{0.2148}\tiny$\pm$0.0113& \underline{0.7924}\tiny$\pm$0.0367& \underline{0.2875}\tiny$\pm$0.0182& \underline{0.8387}\tiny$\pm$0.0238& \underline{0.3128}\tiny$\pm$0.0175& \underline{0.8529}\tiny$\pm$0.0370\\
    & NOCD+DFS& \underline{0.1466}\tiny$\pm$0.0322& \underline{0.7542}\tiny$\pm$0.0598& 0.2046\tiny$\pm$0.0185& 0.7760\tiny$\pm$0.0638& 0.2702\tiny$\pm$0.0158& 0.8261\tiny$\pm$0.0304& 0.3038\tiny$\pm$0.0172& 0.8427\tiny$\pm$0.0178\\
    & \textsf{META-CODE}& {\bf 0.1575}\tiny$\pm$0.0371& {\bf 0.7613}\tiny$\pm$0.0193& {\bf 0.2177}\tiny$\pm$0.0207& {\bf 0.8079}\tiny$\pm$0.0193& {\bf 0.2921}\tiny$\pm$0.0169& {\bf 0.8431}\tiny$\pm$0.0177& {\bf 0.3191}\tiny$\pm$0.0302& {\bf 0.8581}\tiny$\pm$0.0175 \\
    \bottomrule[1pt]
  \end{tabular}
\end{table*}

\textbf{Benchmark methods.} We present three benchmark methods for overlapping community detection: {\bf BIGCLAM} \cite{yang2013overlapping} learns the affiliation matrix using a block coordinate gradient ascent algorithm; {\bf vGraph} \cite{sun2019vgraph} jointly performs community detection and node representation learning based on a generative model; and {\bf NOCD} \cite{shchur2019overlapping} discovers overlapping communities via GNNs using the network structure and node metadata. Since the above benchmark methods were originally developed for {\it fully observable} networks, we need to first infer the network structure. To this end, we employ two sampling strategies for network exploration via node queries: random sampling (RS) and depth-first search (DFS)\footnote{Although other sophisticated sampling strategies such as biased random walks can also be employed, they were not employed in our study since they reveal similar tendencies to those of RS and DFS.}. Similarly as in \textsf{META-CODE}, we iteratively perform community detection alongside the explored network via sampling given a budget of node queries. We implement all these benchmark methods through the parameter settings mentioned in the original articles.

\textbf{Experimental setup.} In our experiments, we adopt GCN \cite{kipf2016semi} with 2 layers as a popular GNN architecture to extract the community-affiliation embeddings. We train our GNN model using Adam optimizer \cite{kingma2014adam} with a learning rate of 0.001. All models are implemented via a geometric learning library, named PyTorch Geometric \cite{fey2019fast}. We tune the hyperparameter $\eta$ within a designated range $\eta\in[1.0, 2.0]$, which leads to substantial gains over the case of $\eta = 0$ corresponding to the GNN model trained in the sense of preserving the structural information only. We also tune the hyperparameter $\lambda$ within a designated range $\lambda \in[0.0, 3.0]$.

\textbf{Research questions.} Our empirical study is designed to answer the following three key research questions (RQs).
\begin{itemize}
\item RQ1. How much does the \textsf{META-CODE} method improve the performance of community detection over the benchmark methods?
\item RQ2. How effective is our initial network inference module for accurate community detection?
\item RQ3. How effective is our node query strategy for fast network exploration?
\end{itemize}

{\bf (RQ1)} \textbf{Comparison with benchmark methods.} The performance comparison between our \textsf{META-CODE} method and three benchmark methods with two different sampling strategies is comprehensively presented in Table \ref{tab:comparison} with respect to two performance metrics using three real-world datasets when different portions of nodes are queried among $N$ nodes. We observe the following: 1) \textsf{META-CODE} consistently outperforms benchmark methods; 2) the performance improvement of \textsf{META-CODE} over the second best performer is the largest when 40\% nodes are queried on Facebook 0---the maximum gain of $23.56\%$ is achieved in terms of the NMI; and 3) the second best performer tends to be NOCD with RS in most cases since NOCD also generates community-affiliation embeddings via GNNs, which is more capable of precisely discovering communities.

\begin{table}[t]
\small
  \caption{Performance comparison of \textsf{META-CODE} with two different initial network inference strategies in terms of the NMI when 10\% of nodes are queried.}
  \begin{tabular}{cccl}
    \toprule
    Dataset&\textsf{META-CODE} ($k$NN)&\textsf{META-CODE} (MAC+AGM)\\
    \midrule
    Facebook 0 & 0.0272\tiny$\pm$0.0194& {\bf 0.0298}\tiny$\pm$0.0157 \\
    Facebook 348& 0.0772\tiny$\pm$0.0174& {\bf 0.0791}\tiny$\pm$0.0198 \\
    Engineering& 0.1431\tiny$\pm$0.0204 & {\bf 0.1575}\tiny$\pm$0.0371 \\
  \bottomrule
\end{tabular}
\label{table:NetworkInference}
\end{table}

{\bf (RQ2)} \textbf{Effectiveness of initial network inference.} We investigate the effect of our initial network inference module based on the MAC and AGM on the performance of \textsf{META-CODE} in comparison with the case of using a $k$-nearest neighbor ($k$NN)-aided strategy. The performance comparison between two different strategies is presented in Table \ref{table:NetworkInference} with respect to the NMI when 10\% of node are queried. Our observations include that 1) \textsf{META-CODE} (MAC+AGM) is consistently superior to \textsf{META-CODE} ($k$NN) and 2) the maximum gain of 9.55\% is achieved on Facebook 0.

\begin{figure}[t]
\begin{center}
    \includegraphics[width=1.0\linewidth]{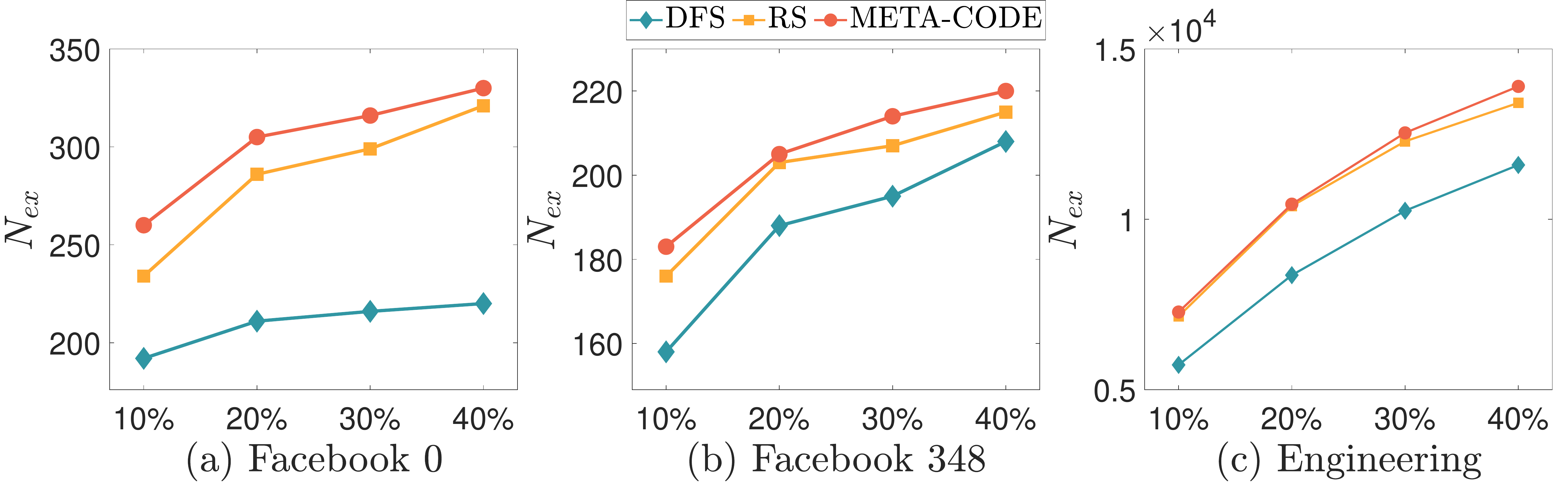}
\end{center}
   \caption{The number of explored nodes, $N_\text{ex}$, according to different portions (\%) of queried nodes among $N$ nodes.}
\label{fig:exploration}
\end{figure}

{\bf (RQ3)} \textbf{Growth of explored subgraphs.} To empirically validate the effectiveness of our node query strategy, we conduct experiments by plotting the number of explored nodes (i.e., the size of explored subgraphs), denoted as $N_\text{ex}$,  versus a portion of queried nodes in Figure \ref{fig:exploration} in comparison with two cases in which two sampling strategies (i.e., RS and DFS) are employed for network exploration in \textsf{META-CODE} rather than Eq. (\ref{eq:strategy}). Our findings are summarized as follows: 1) our query strategy based on the discovered community affiliations is consistently beneficial over other samplings with respect to the growth of explored subgraphs by exploring the unknown parts of the underlying network faster; and 2) the growth rate of $N_\text{ex}$ for RS is even higher than that for DFS since querying nodes away from explored subgraphs via RS (instead of expanding the explored subgraph via DFS) is more effective as the number of queried nodes increases.

\section{Concluding Remarks}

We introduced \textsf{META-CODE}, a novel end-to-end solution for overlapping community detection in networks with unknown topology. Our \textsf{META-CODE} method was designed in such a way of leveraging limited node queries and node metadata to iteratively generate GNN-based embeddings and identify queried nodes after initial network inference. Using three real-world datasets, we demonstrated (a) the superiority of \textsf{META-CODE} over competitors with substantial gains up to $23.56\%$ and (b) the effectiveness of key components in \textsf{META-CODE} such as the initial network inference module and the network exploration module via node queries.

\begin{acks}
This work was supported by the National Research Foundation of Korea (NRF) grant funded by the Korea government (MSIT) (No. 2021R1A2C3004345).
\end{acks}

\bibliographystyle{ACM-Reference-Format}
\balance
\bibliography{sample-base}


\end{document}